# Designing New Electrolytes for Lithium Ion Batteries Using Superhalogen Anions

Ambrish Kumar Srivastava and Neeraj Misra*

Department of Physics, University of Lucknow, Lucknow - 226007, Uttar Pradesh, India

*Corresponding author, E-mail: neerajmisra11@gmail.com

# Abstract

The electrolytes used in Lithium Ion Batteries (LIBs) such as $LiBF_4$, $LiPF_6$ etc. are Li-salts of some complex anions, $BF_4^-$, $PF_6^-$ etc. The investigation shows the vertical detachment energy (VDE) of these anions exceeds to that of halogen, and therefore they behave as superhalogen anions. Consequently, it might be possible to design new electrolytic salts using other superhalogen anions. We have explored this possibility using Li-salts of various superhalogen anions such as $BO_2^-$, $AlH_4^-$, $TiH_5^-$ and $VH_6^-$ as well as hyperhalogen anions, $BH_{4-y}(BH_4)_y^-$ (y = 1−4). Our density functional calculations show that Li-salts of these complex anions possess similar characteristics as those of electrolytic salts in LIBs. Note that they all are halogen free and hence, non-toxic and safer than $LiBF_4$, $LiPF_6$ etc. In particular, $LiB_4H_{13}$ and $LiB_5H_{16}$ are two potential candidates for electrolytic salt due to their smaller Li-dissociation energy ($\Delta E$) than those of $LiBF_4$, $LiPF_6$ etc. We have also noticed that $\Delta E$ of $LiBH_{4-y}(BH_4)_y$ varies inversely with the VDE of $BH_{4-y}(BH_4)_y^-$ anions, which increases with the increase in y. These findings may guide experimentalists and future researchers to design and synthesize more efficient and environment friendly electrolytic salts for LIBs.

**Keywords:** Superhalogens; Lithium salts; Electrolytes; Lithium Ion Battery; Density functional calculations.

## 1. Introduction

Lithium-ion batteries (LIBs) have become the main power source of modern electronic devices due to their high density of energy storage [1, 2]. The multifunctional cellular phones and expanding applications of LIBs in other power storage systems demand a persistent increase in its energy storage capacity. This necessitates the development of potentially active materials, the task which has been attracting attention of experimentalists as well as theoreticians [3, 4]. A typical LIB consists of a graphite anode and transition metal oxide complex of Li (such as $LiCoO_2$, $LiMn_2O_2$ etc.) as cathode. These two electrodes are separated by a non-aqueous organic electrolyte that acts as an ionic medium. The electrolytes are made up of non-coordinating anionic complexes of Li, such as $LiBF_4$, $LiPF_6$, $LiClO_4$ and $LiFePO_4$ *etc* dissolved in a polar aprotic solvent. It is generally accepted that organic electrolytes are decomposed during the first lithium intercalation into graphite to form a solid electrolyte interface (SEI) film between the graphite anode surface and the electrolyte, and it is the SEI film that largely determines the performance of graphite as anode in rechargeable batteries [1]. For instance, the higher the film passivating ability, the better capacity and longer life cycle of the lithium-intercalated graphite anodes. A number of efforts have been made on the investigations of various organic solvents in order to increase the reversibility and capacity of LIBs [5, 6]. Most common organic solvents include esters or ethers or their mixtures. In this line, ethylene carbonate and propylene carbonates have extensively been studied by Wang et al. [7, 8] and they have theoretically suggested a variety of additives to these solvents, which are beautifully reviewed by Zhang [9]. Most of the currently used electrolytic salts such as $LiBF_4$, $LiPF_6$, $LiClO_4$, etc. contain halogen and hence, they can not be considered as safe from environmental perspectives. Therefore, it is desirable to quest for new electrolytic salts, which are safe and at the same time, as efficient (conducting) as existing salts in LIBs.

A recent investigation [10] has indicated the superhalogen as building block of electrolytes in LIBs. The word ‘superhalogen’ refers to the species, whose electron affinity exceed to that of halogen atoms. Being proposed by Gutsev and Bolyrev [11, 12] for the first time during 1980s, these species have attracted continuous attentions [13-20]. Such species possess strong oxidizing property [21] and form ionic complexes with appropriate metal atoms [18, 19]. A general formula of $MX_{k+1}$ has been proposed [11] for typical superhalogen species, where M is a core atom with formal valence of *k* and X is electronegative ligands such as F, Cl, O, CN, OF etc. The electrolytic salts in LIBs are lithium complexes of superhalogen anions. For instance, $LiBF_4$ and $LiPF_6$ can be thought of as $Li^+BF_4^-$ and $Li^+PF_6^-$, respectively, where $BF_4$ and $PF_6$ are superhalogens satisfying $MX_{k+1}$ formula. A recent study [22] has demonstrated that the electronic properties of $LiBeF_3$ closely resemble those of $LiBF_4$ and $LiPF_6$. Therefore, it is expected that the concept of superhalogen might be useful in designing new electrolytic salts in LIBs, as noticed in previous investigation [10], which proposed Li-salts of $BeF_3^-$, $AuF_6^-$, $NO_3^-$, $BH_4^-$, $B_3H_8^-$ and $CB_{11}H_{12}^-$ superhalogen anions as a possible candidate for electrolytes in LIBs. Note that $LiBeF_3$ and $LiAuF_6$ do contain halogen and hence, are not as safe as expected. Furthermore, $LiNO_3$ is sufficiently conducting to be used as electrolytic salt. This prompted us to design new electrolytes using the concept of superhalogen, which might be more efficient than proposed salts in the previous study [10].

## 2. Computational details

All lithium salts (Li−X) and corresponding anions ($X^-$) considered in this study were fully optimized at ωB97xD method [23] using 6-311+G(d) basis set [24] using Gaussian 09 program [25]. The functional ωB97xD includes empirical dispersion, which has been found appropriate for the systems with long range interactions [26]. Furthermore, the present computational scheme, ωB97xD/6-311+G(d), has already been used in the previous study on Li-salts [10]. The geometry optimization was performed without any symmetry constraints and followed by frequency calculations to ensure that the optimized structures correspond to true minima in the potential energy surface. The vertical detachment energy (VDE) of anions has been calculated by difference of total energy of optimized structure of anions and corresponding neutral structure, i.e.,

$$\text{VDE} = E[\text{X}]_{\text{single point}} - E[\text{X}^-]_{\text{optimized}}$$

where $E[X^-]_{optimized}$ is the total energy of optimized structure of $X^-$ and $E[X]_{single\ point}$ is the single point energy of neutral structure (X) at optimized geometry of $X^-$.

## 3. Results and discussion

We start our discussion with typical electrolytic salts in LIBs, which include $LiBF_4$, $LiPF_6$, $LiClO_4$, $LiFePO_4$ and $LiAsF_6$ etc. The VDE of corresponding anions are listed in Table 1. One can note that all VDE values exceed the electron affinity of halogen atom, which is limited to 3.59 eV for Cl [27]. Therefore, these anions belong to the class of superhalogen. The optimized structure of $BF_4^-$, $PF_6^-$ and their Li-salts are displayed in Fig. 1. Since most of these anions consist of halogen atoms, which are toxic and therefore environmental safety remains a grave issue. As an attempt to overcome this issue, Li-salts of $BeF_3^-$, $AuF_6^-$, $NO_3^-$, $BH_4^-$, $B_3H_8^-$ and $CB_{11}H_{12}^-$ have been considered [10]. The VDE of these anions are also listed in Table 1, which establish their superhalogen behavior. Note that $BeF_3^-$ and their compounds are poisonous due to toxicity of beryllium content and $AuF_6^-$ does contain halogen. In fact, many transition metal hexafluorides have been reported to possess higher electron affinity than halogen, excluding $WF_6$ [28]. Therefore, they all possess more or less similar properties and form almost similar Li-salts, like $LiAuF_6$, $LiAsF_6$ and $LiPF_6$. However, they all suffer with certain level of toxicity due to presence of halogen. The salt of $CB_{11}H_{12}^-$ has been found to possess more favourable properties [10] and hence, $LiCB_{11}H_{12}$ has been advocated as a possible candidate for electrolytes in future LIBs. However, it contains carbon unlike most of the electrolytic salts, which are purely inorganic compounds. Therefore, a rigorous search for superhalogen anions, whose Li-salts could be employed as electrolytes in LIBs, is needed.

In Fig. 2, we have displayed the structures of some anions such as $BO_2^-$, $AlH_4^-$, $TiH_5^-$ and $VH_6^-$. The VDEs of these anions are also listed in Table 1, which suggest that these species also belong to the superhalogen series. One of the advantages of these species is that they all are halogen free and hence, not toxic. The equilibrium structures of Li-salts of these anions are also displayed in Fig. 2. We have also designed some borohydride $BH_{4-y}(BH_4)_y^-$ (y = 1−4) anions by successive replacement of H atoms by $BH_4^-$ moieties in $BH_4^-$ superhalogen. These are a special class of superhalogens, referred to as hyperhalogens [29, 30], in which the role of ligands is played by a superhalogen itself. The equilibrium structures of resulting $B_2H_7^-$, $B_3H_{10}^-$, $B_4H_{13}^-$ and $B_5H_{16}^-$ hyperhalogens are displayed in Fig. 3 along with their Li-salts. The VDE of these hyperhalogens increases successively with the increase in $BH_4$ ligands and reaches to as high as 7.28 eV for $B_5H_{16}^-$.

In order to behave as electrolytes, Li-salts should add to the conductivity of electrolytic medium. It is known that electrolytic salt contributes a lithium ion by dissociating into ionic fragments. Therefore, we have considered the dissociation of Li−X complexes into $Li^+$ and $X^-$. The corresponding dissociation energy ($\Delta E[Li^+]$) is calculated as follows:

$$\Delta E[Li^+] = E[Li^+] + E[X^-] - E[Li-X]$$

where $E$[..] represent total electronic energy of respective species including zero point correction. For an efficient electrolyte, $\Delta E[Li^+]$ should be comparable to or smaller than those for commercial electrolytes. The calculated $\Delta E[Li^+]$ values for various Li-salts are listed in Table 2. One can note that $\Delta E[Li^+]$ of $LiPF_6$ (5.73 eV) is smaller than that of $LiBF_4$ (6.08 eV) and therefore, it should be more conducting. The $\Delta E[Li^+]$ of previously proposed electrolytic salts is 5.08 eV for $LiCB_{11}H_{12}$ and 5.50 eV for $LiAuF_6$. Furthermore, $\Delta E[Li^+]$ of other proposed salts ranges from 6.25 eV ($LiB_3H_8$) to 6.62 eV ($LiBH_4$) [10]. For our proposed salts, $\Delta E[Li^+]$ varies from 6.00 eV ($LiAlH_4$) to 6.36 eV ($LiBO_2$), which are comparable to those of the commercial electrolytic salts. Note that $LiBH_4$ has already been experimentally studied for its potential application as electrolytic salt [31]. Recently, stable lifecycle of all solid state LIB employing pure $LiBH_4$ solid electrolyte with $LiCoO_2$ and Li metal as cathode and anode, respectively, has been reported [32]. $LiAlH_4$ seems to be a better candidate than $LiBH_4$ due to smaller $\Delta E[Li^+]$ value and hence, increased conductivity. The $\Delta E[Li^+]$ of $LiBH_{4-y}(BH_4)_y$ (y = 1−4) complex salts are even more interesting. For example, $\Delta E[Li^+]$ of $LiB_2H_7$ is smaller than $LiBH_4$, which further decreases for $LiB_3H_{10}$. Moreover, $\Delta E[Li^+]$ values of $LiB_4H_{13}$ and $LiB_5H_{16}$ are smaller than those of $LiBF_6$ and $LiPF_6$, respectively (see Table 2). We have also noticed an interesting trend between VDE of borohydride hyperhalogen anions and $\Delta E[Li^+]$ of their Li-complexes, as plotted in Fig. 4. One can see that $\Delta E[Li^+]$ decreases successively with the increase in $BH_4$ moieties ($y$) in $BH_{4-y}(BH_4)_y^-$ (y = 1−4) hyperhalogen anions. Thus, $\Delta E[Li^+]$ of $LiBH_{4-y}(BH_4)_y$ salts depends inversely on their VDEs. In the light of $\Delta E[Li^+]$ values, $LiB_4H_{13}$ and $LiB_5H_{16}$ are two potential candidates for electrolytic salt in LIBs. Furthermore, larger size of $LiB_4H_{13}$ and $LiB_5H_{16}$ offer greater mobility of $Li^+$ ions than $LIBH_4$ due to larger size of complex anions.

It is also required for an electrolytic salt to be sufficiently stable. In order to analyze the stability of Li-salts against complexation with water ($H_2O$), we have calculated the complexation energy $\Delta E[H_2O]$ as below:

$$\Delta E[H_2O]= E[Li-X] + E[H_2O] - E[H_2O-Li-X]$$

The optimized structures of $H_2O$ complexes of Li−X salts are can be obtained in supplementary information. It is desirable that the water affinity of salts should be comparable to or smaller than that of traditional electrolytic salts, i.e., $\Delta E[H_2O]$ of salts should be comparable to or greater than those of $LiBF_4$ and $LiPF_6$. The calculated $\Delta E[H_2O]$ of Li-salts are also listed in Table 2, which suggest that proposed salts are indeed stable. The $\Delta E[H_2O]$ values of $LiBH_{4-y}(BH_4)_y$ (y = 1−4) are also equal to that of $LiBH_4$ and $LiB_3H_8$ proposed earlier. Therefore, these Li-borohydride complexes possess more favourable characteristics to be employed as electrolytic salts. Note that many efforts have already been made to stabilize $LiPF_6$ by choosing appropriate additives to electrolytic solvents [9]. Likewise, the proposed salts can be further stabilized by modifying the electrolytic solvents. Now we compare the properties of our proposed salts with those of Giri et al. [10]. Our proposed salts are halogen free inorganic species, so they are preferable from environmental safety point of view. Furthermore, some of these salts possess quite adequate properties to qualify as an electrolytic salt. For instance, $LiBO_2$ possesses lower $\Delta E[Li^+]$ and higher $\Delta E[H_2O]$ values than those of $LiNO_3$. Similarly, $\Delta E[Li^+]$ and $\Delta E[H_2O]$ values of $LiAlH_4$ make it more favourable as electrolytic salts than $LiBH_4$ proposed earlier. $LiBH_{4-y}(BH_4)_y$ (y = 1−4) proposed here also show its potential as electrolytic salt, whose $\Delta E[Li^+]$ value is smaller than that of $LiBH_4$ and $LiB_3H_8$ but $\Delta E[H_2O]$ values are almost equal. Moreover, our proposed $LiTiH_5$ and $LiVH_6$ complexes also possess similar characteristics to be utilized as electrolytic salts in LIBs.

## 4. Conclusions

Using concept of superhalogen, we have reinvestigated the possibility of design of new electrolytic salts for LIBs. We have considered Li-salts of $BO_2^-$, $AlH_4^-$, $TiH_5^-$ and $VH_6^-$ superhalogen anions as well as a series of borohydride $BH_{4-y}(BH_4)_y^-$ (y = 1−4) hyperhalogen anions, which all are halogen free and hence, more environment friendly. Their ionization to $Li^+$ requires the energy comparable to commercial electrolytic salts such as $LiBF_4$. Therefore, the performance of LIBs using these electrolytic salts is not much affected. Their water affinity is also comparable to traditional electrolytic salts such as $LiPF_6$ and hence, the life cycle of LIBs is also preserved using these halogen free electrolytic salts. Thus, the superhalogen anions are indeed useful in designing new electrolytic salts for LIBs. In particular, $LiB_4H_{13}$ and $LiB_5H_{16}$ are found to possess the most desirable characteristics for electrolytic salts such as lower $Li^+$ ionization energy and greater mobility of $Li^+$ ions due to larger size of complex anions. These finding might be useful for designing more efficient electrolytes for future LIBs.

## Supplementary Information

The optimized structures of $H_2O$ complexes of Li-salts considered in this study are available.

## Acknowledgement

The author, A. K. Srivastava acknowledges Council of Scientific and Industrial Research (CSIR), New Delhi, India for a research fellowship [Grant No. 09/107(0359)/2012-EMR-I].

Table 1. The VDE of complex anions for electrolytic salts at ωB97xD/6-311+G(d) level.

| Currently used | VDE (eV) | Proposed[a] | VDE (eV) | Proposed[b] | VDE (eV) |
|---|---|---|---|---|---|
| $FePO_4$ | 4.32 | $NO_3$ | 4.22 | $VH_6$ | 4.23 |
| $ClO_4$ | 5.83 | $BH_4$ | 4.50 | $BO_2$ | 4.44 |
| $N(SO_2F)_2$ | 6.89 | $B_3H_8$ | 4.72 | $AlH_4$ | 4.69 |
| $N(SO_2CF_3)_2$ | 7.01 | $CB_{11}H_{12}$ | 5.99 | $TiH_5$ | 4.72 |
| $BF_4$ | 7.66 | $BeF_3$ | 6.99 | $B_2H_7$ | 5.52 |
| $PF_6$ | 8.55 | $AuF_6$ | 8.66 | $B_3H_{10}$ | 6.25 |
| $AsF_6$ | 8.91 | | | $B_4H_{13}$ | 6.64 |
| | | | | $B_5H_{16}$ | 7.28 |

[a]Ref. [10]

[b]This work

Table 2. Li-ionization energy ($\Delta E[Li^+]$) and water complexation energy ($\Delta E[H_2O]$) of various Li-salts (in eV) calculated at ωB97xD/6-311+G(d) level.

| Salt | $\Delta E[Li^+]$ | $\Delta E[H_2O]$ |
|---|---|---|
| $LiBF_4$ | 6.08 | 1.41 |
| $LiPF_6$ | 5.73 | 1.07 |
| $LiNO_3$[a] | 6.53 | 0.96 |
| $LiBH_4$[a] | 6.62 | 0.92 |
| $LiB_3H_8$[a] | 6.25 | 0.93 |
| $LiCB_{11}H_{12}$[a] | 5.08 | 1.08 |
| $LiAuF_6$[a] | 5.50 | 1.06 |
| $LiVH_6$ | 6.29 | 0.91 |
| $LiBO_2$ | 6.36 | 0.99 |
| $LiAlH_4$ | 6.00 | 0.97 |
| $LiTiH_5$ | 6.11 | 0.90 |
| $LiB_2H_7$ | 6.34 | 0.92 |
| $LiB_3H_{10}$ | 6.15 | 0.91 |
| $LiB_4H_{13}$ | 5.80 | 0.91 |
| $LiB_5H_{16}$ | 5.45 | 0.93 |

[a]Ref. [10]

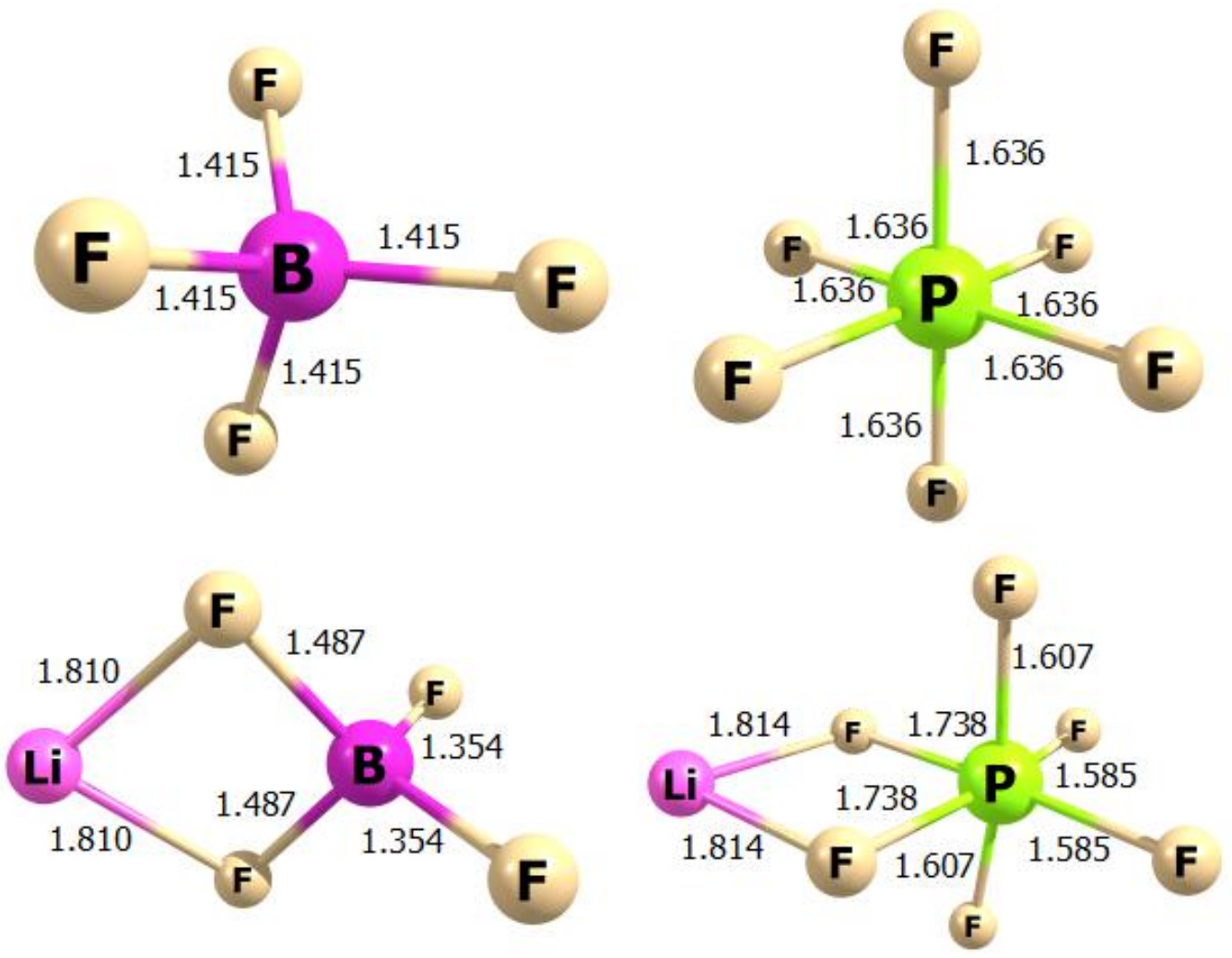

Fig. 1. Equilibrium structures of $BF_4^-$, $PF_6^-$ superhalogen anions and their Li-salts, which are currently used as electrolytic salts in LiBs. The bond lengths in Å are calculated at ωB97xD/6-311+G(d) level.

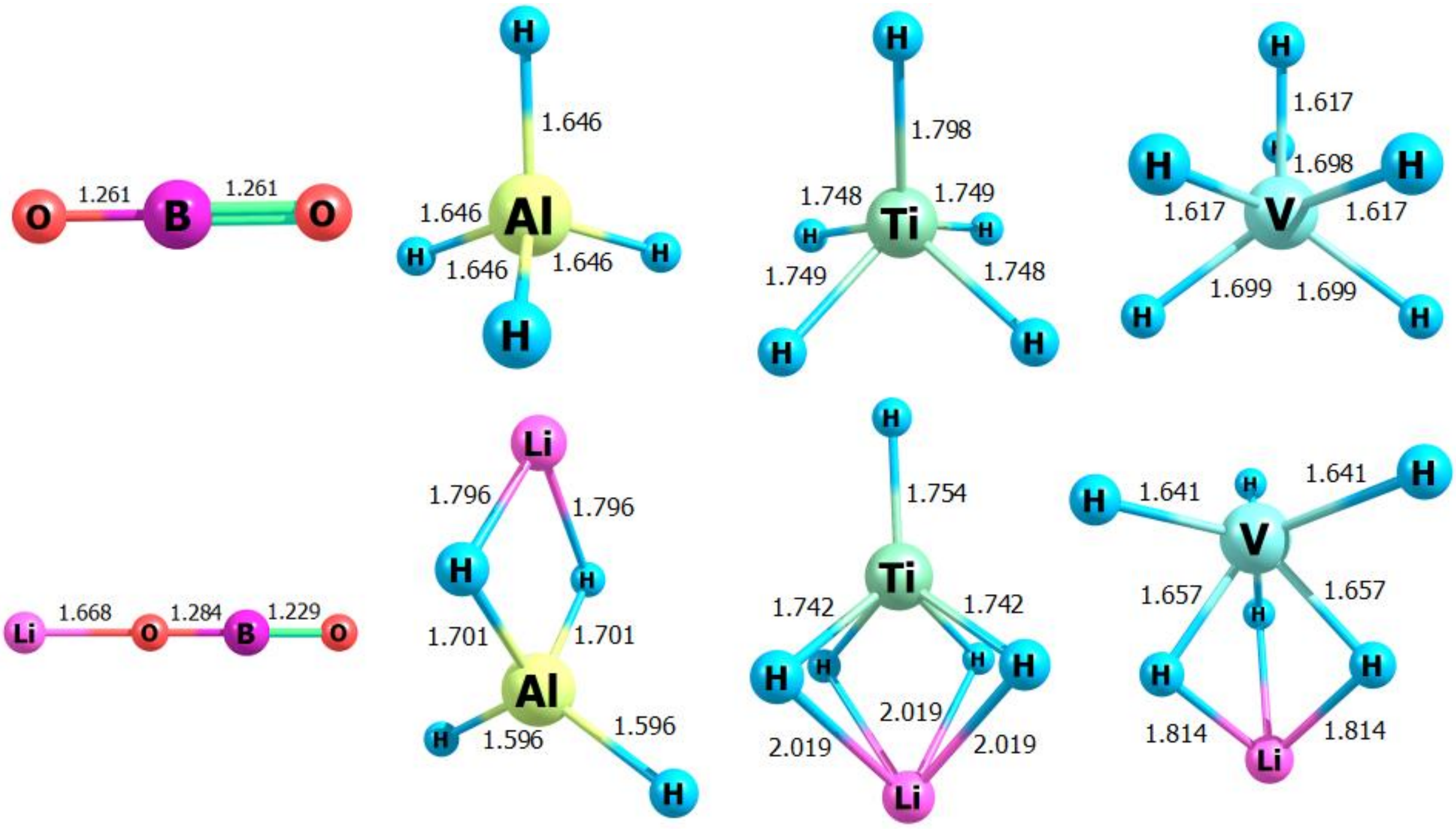

Fig. 2. Equilibrium structures of $BO_2^-$, $AlH_4^-$, $TiH_5^-$ and $VH_6^-$ superhalogen anions along with their Li-salts. The bond lengths in Å are calculated at ωB97xD/6-311+G(d) level.

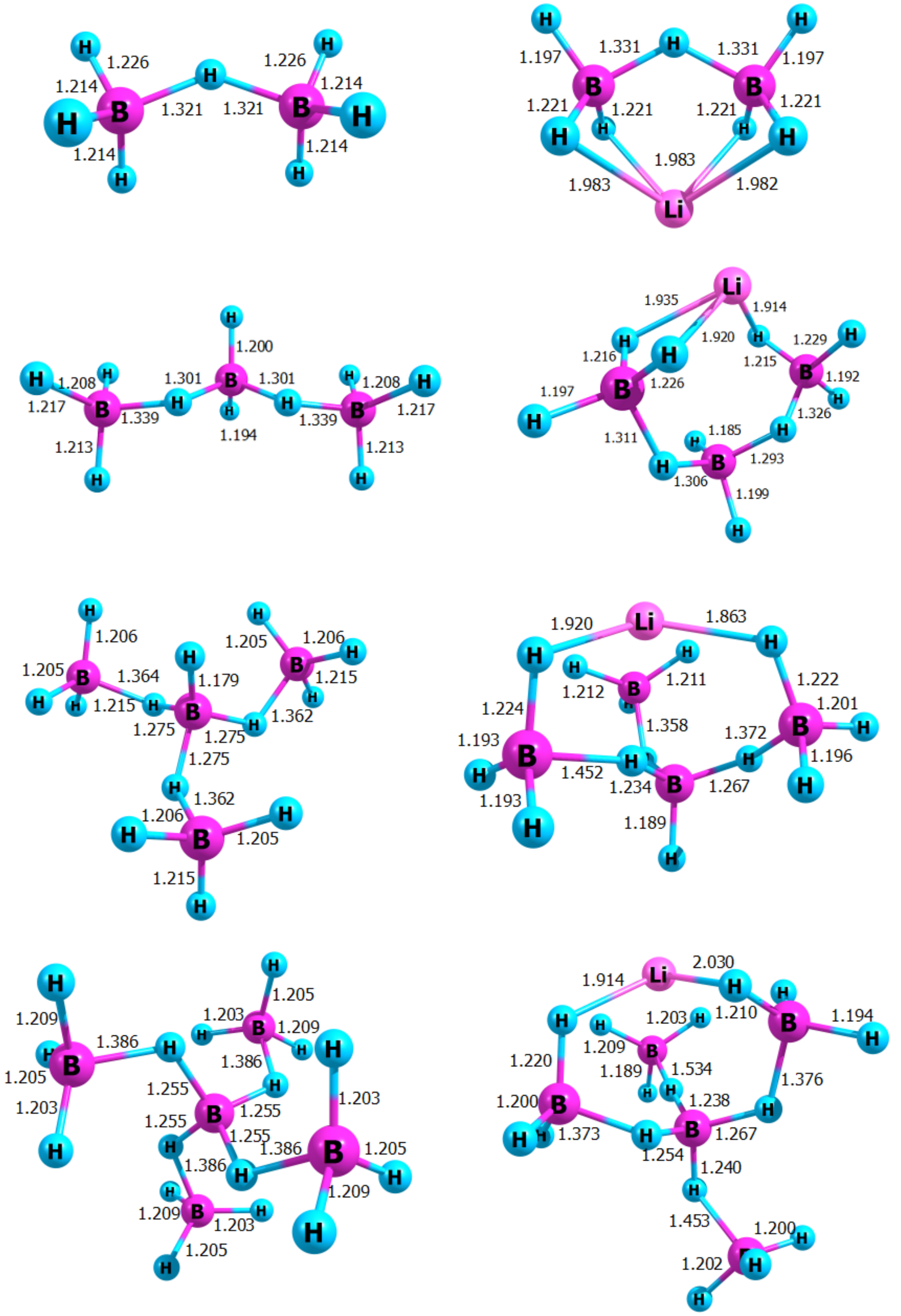

Fig. 3. Equilibrium structures of $BH_{4-y}(BH_4)_y^-$ (y = 1−4) hyperhalogen anions and their Li-salts. The bond lengths in Å are calculated at ωB97xD/6-311+G(d) level.

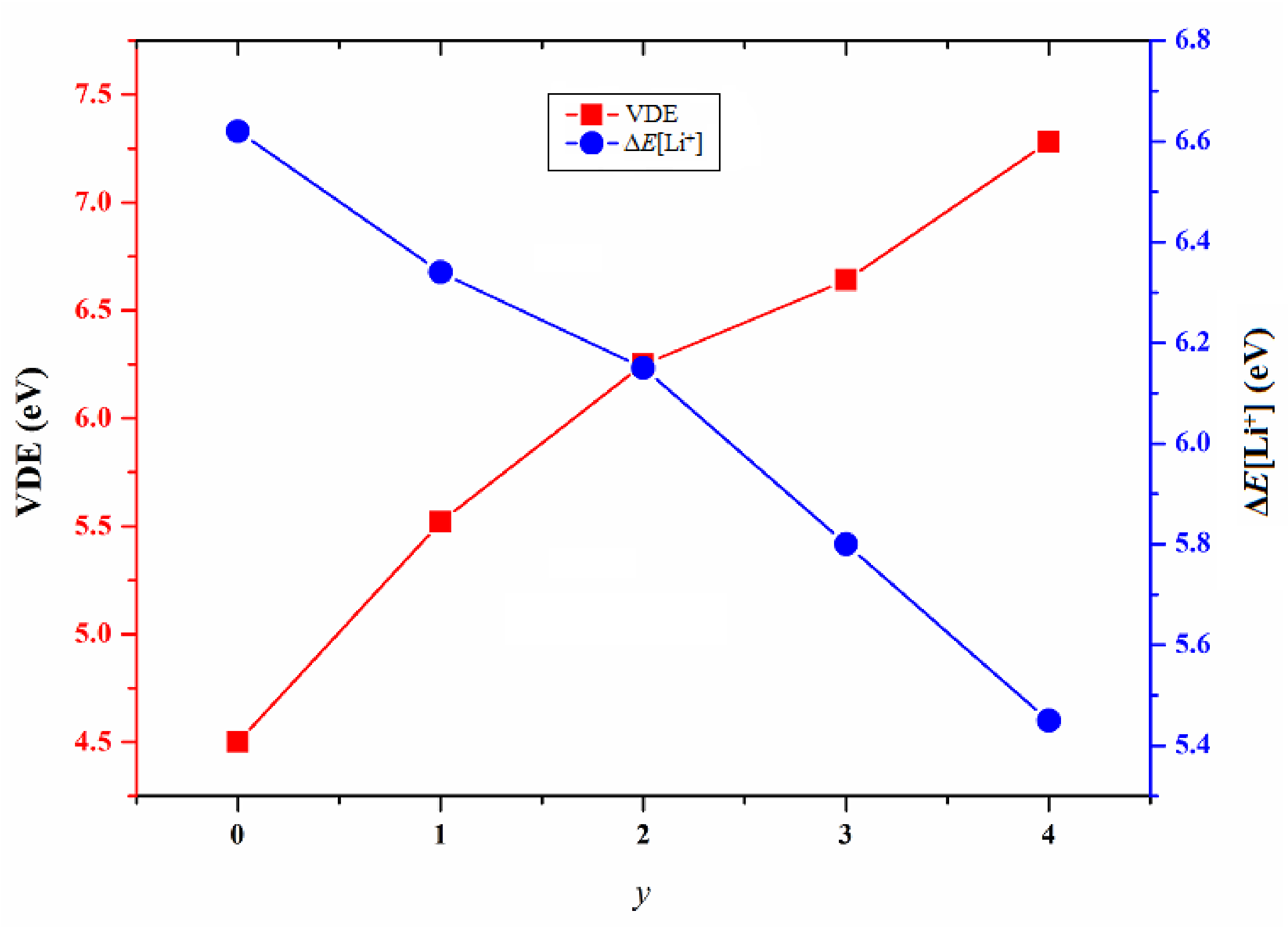

Fig. 4. The plots of VDE of $BH_{4-y}(BH_4)_y^-$ hyperhalogen anions and $\Delta E[Li^+]$ of $LiBH_{4-y}(BH_4)_y$ salts as a function of $y$.